\documentclass[preprint]{aastex}
\usepackage{natbib}
\usepackage{graphicx}
\usepackage{epstopdf}

\def\asec{\ifmmode ^{\prime\prime}\else$^{\prime\prime}$\fi}

\def\it{\sl}
\def\degs{\ifmmode ^{\circ}\else$^{\circ}$\fi}
\def\amin{\ifmmode ^{\prime}\else$^{\prime}$\fi}
\def\asec{\ifmmode ^{\prime\prime}\else$^{\prime\prime}$\fi}
\def\fm{\hbox{$.\!\!^{\rm m}$}}            % Fractions of magnitudes
\def\farcs{\hbox{$.\!\!^{\prime\prime}$}}  % Fractions of arcseconds
\def\degs{\ifmmode ^{\circ}\else$^{\circ}$\fi}
\def\amin{\ifmmode ^{\prime}\else$^{\prime}$\fi}
\def\farcm{\hbox{$.\mkern-4mu^\prime$}}
\def\eqalign#1{\null\,\vcenter{\openup1\jot \m@th
   \ialign{\strut\hfil$\displaystyle{##}$&$\displaystyle{{}##}$\hfil
   \crcr#1\crcr}}\,}

\slugcomment{to appear in The Astrophysical Journal}

\begin{document}


\title{Deep optical observations of unusual neutron star Calvera with the GTC\altaffilmark{\dag}}


\author{Yury Shibanov\altaffilmark{1,2}, Andrey Danilenko\altaffilmark{1}, Sergey Zharikov\altaffilmark{3}, Peter Shternin\altaffilmark{1}, Dima Zyuzin\altaffilmark{1}}

\altaffiltext{1}{Ioffe Institute, Politekhnicheskaya 26, St.~Petersburg, 194021, Russia}
\altaffiltext{2}{Peter The Great St. Petersburg Polytechnic University, Politekhnicheskaya 29, St.~Petersburg, 195251, Russia}
\altaffiltext{3}{Instituto de Astronom\'{\i}a, Universidad Nacional Aut\'{o}noma de M\'{e}xico, Apartado Postal 877, Ensenada, Baja California, 22800 M\'{e}xico}

\altaffiltext{\dag}{Based on observations made with the Gran Telescopio Canarias (GTC), instaled in the Spanish Observatorio del Roque de los Muchachos of the Instituto de Astrofísica de Canarias, in the island of La Palma, programme GTC1-14AMEX.}



\begin{abstract}
Calvera is an unusual isolated neutron star with pure thermal X-ray spectrum typical for central compact objects in supernova remnants. On the other hand, its rotation period and spin-down rate are typical for ordinary rotation-powered pulsars. It was discovered and studied  in X-rays and not yet detected in other spectral domains. We present deep optical imaging of the Calvera field obtained with the Gran Telescopio Canarias in $g'$ and $i'$ bands. Within $\approx 1\asec$ vicinity of Calvera, we detected two point-like objects invisible at previous shallow observations. However, accurate astrometry showed that none of them can be identified with the pulsar. We put new upper limits on its optical brightness of $g' > 27.87$ and $i' > 26.84$. We also reanalyzed all available archival X-ray data on Calvera. Comparison of the Calvera thermal emission parameters and upper limits on optical and non-thermal X-ray emission with respective data on rotation-powered pulsars shows that Calvera might belong to the class of ordinary middle-aged pulsars, if we assume that its distance is in the range of 1.5--5 kpc.
\end{abstract}


\section{Introduction} \label{sec:intro}

High Galactic latitude, $b \approx +37\degs$, compact source 1RXS~J141256.0+792204 in the \textit{ROSAT} All-Sky Survey was proposed as a neutron star (NS) candidate  by \citet{rutledge2008ApJ} based on its high X-ray to optical flux ratio. It was the first plausible isolated NS (INS) candidate found in the Survey after the identification of 7 purely thermally emitting radio-quiet INSs dubbed ``The Magnificent Seven''. This led \citet{rutledge2008ApJ} to nickname 1RXS~J141256.0+792204 as ``Calvera'', another character of the same fiction movie. Subsequent observations confirmed the NS nature of Calvera and revealed unusual properties making it difficult to attribute the source to any specific class of NSs \citep[for details see][]{zane2011MNRAS,halpern2013ApJ}.

The initial idea that Calvera can be a new member of the Magnificent Seven family was ruled out by X-ray timing observations. Using \textit{XMM-Newton}, \citet{zane2011MNRAS} discovered  X-ray pulsations with the period $P\approx59$~ms. \textit{Chandra} observations then allowed \citet{halpern2013ApJ} to measure and later to refine \citep{Halpern2015} the spin-down rate with the period derivative $\dot{P} \approx 3.2\times  10^{-15}$~s~s$^{-1}$, yielding the spin-down luminosity $\dot{E} \approx 6.1\times 10^{35}$ erg~s$^{-1}$, the characteristic age $\tau \approx 290$ kyr, and the dipole magnetic field $B \approx 4.4\times 10^{11}$ G. These values are drastically different from those of Magnificent Seven NSs which are much less energetic ($\dot{E}\sim 10^{30}-10^{32}$ erg~s$^{-1}$), slowly-rotating ($P\sim 1-10$~s), old ($\tau\sim 1$~Myr), and highly-magnetized ($B\sim 10^{13}$~G) INSs \citep[e.g.,][]{haberl2007,kaplan2011ApJ}.

Alternative interpretation of Calvera as an ordinary rotation-powered pulsar (RPP) is also challenging. Deep searches failed to detect it in the radio \citep{hessels2007,zane2011MNRAS}, and no evident non-thermal emission component expected for such an energetic pulsar was found in X-rays \citep{zane2011MNRAS,halpern2013ApJ}. Neither steady nor pulsed signals were detected with \textit{Fermi} in $\gamma$-rays \citep{halpern2011ApJ, halpern2013ApJ}. The out-of-beam scenario is frequently considered for the non-detection of a narrow-beamed pulsar radio emission. However, it is hardly plausible in the high energy bands where pulsar beams are much wider. The non-thermal emission component can be suppressed for aligned magnetic rotators. But Calvera shows a high X-ray pulsed fraction of $\approx 18\%$ \citep{zane2011MNRAS} excluding this possibility. It was also speculated that Calvera can be a descendant from one of central compact objects (CCOs), weakly magnetized ($<10^{11}$ G) thermally emitting NSs observed in supernova remnants, whose magnetic field was initially buried in the NS crust by a prompt fall-back of the parent supernova ejecta and now is emerging back to the surface \citep{gotthelf2013,Halpern2015}.

The proper motion of Calvera, $\mu=69\pm26$~mas~yr$^{-1}$, was recently measured with \textit{Chandra} by \citet{Halpern2015} with the proper motion vector pointing outside of the Galactic plane. The authors argue that Calvera could have been born inside the Galactic disk and now is at a distance $\lesssim 0.3$ kpc with transverse velocity $\la120$ km s$^{-1}$. However, they did not find any suitable birthplace of Calvera within the disk. Moreover, if Calvera is at a distance $\lesssim 0.3$ kpc, it is  by several orders of magnitude under-luminous in $\gamma$-rays, as compared to other pulsars with similar $\dot{E}$ \citep{halpern2013ApJ}. On the other hand, as it was noted by \citet{halpern2013ApJ}, Calvera may be much more distant if it was born in the Galactic halo or its progenitor was a runaway star.

Thus, despite the intensive radio, X-ray, and $\gamma$-ray studies, Calvera remains a puzzling NS.  Its distance, classification, and evolution status are still uncertain and need further multiwavelength  studies. The high Galactic latitude and hence low absorption column density make it a promising target for optical observations. Until now, the Calvera field was observed with the Gemini-N telescope in the $g$ band but no optical counterpart of Calvera was found down to a brightness limit $g \ga 26.3$ \citep{rutledge2008ApJ}.

We report new much deeper optical imaging of the Calvera field in $g'$ and $i'$ bands performed with the Gran Telescopio Canarias (GTC). We discovered a faint red object ($g' = 26.22$ and $i' = 24.17$) located $\sim 0\farcs7$ of the Calvera position. Accurate astrometry of the optical and X-ray images shows that the object can hardly be an optical counterpart of Calvera. We put upper limits on the Calvera optical fluxes. We also re-analyze all the available X-ray data and argue that the thermal X-ray spectrum of Calvera can be described by the magnetized hydrogen atmosphere model  with either uniform or non-uniform temperature distribution corresponding to the emission from the bulk of NS surface located at a distance of 1.5--5~kpc \citep[NSMAX model;][]{ho2008ApJS}. We also confirm previous claims of the presence of a spectral absorption feature in the X-ray spectra of Calvera. Overview of the optical observations, data reduction, and archival data used in the analysis is presented in Section~\ref{sec:data}. Analysis of the optical data and searching for an optical counterpart of Calvera are given in Sections~\ref{sec:astrometry}--\ref{sec:phot-anal}. Spectral analysis of the X-ray data is reported in Section~\ref{sec:spect}. We discuss implications of our results in Section~\ref{sec:disc}.

\section{Observations of Calvera and archival data} \label{sec:data}

\subsection{GTC observations and data reduction and calibration} \label{sec:GTC-data}

The observations of the pulsar field were carried out in the Sloan $g'$ and $i'$ bands with the Optical System for Imaging and low-intermediate Resolution Integrated Spectroscopy (OSIRIS\footnote{http://www.gtc.iac.es/instruments/osiris/}) at the GTC in a queue-scheduled service mode in 2014 April and June. With the image scale of 0\farcs254 pixel$^{-1}$ ($2\times 2$ binning) and unvignetted field size of $7\farcm8\times 7\farcm8$ available with the OSIRIS detector consisting of a mosaic of two CCDs, we obtained three sets of  dithered exposures. Calvera was exposed on  CCD2. The observing conditions were clear with stellar flux variations of less than 4\% and with seeing varied from 0\farcs7 to 1\farcs1 (see Table~\ref{t:log}).

Standard data reduction including bias subtraction, flat-fielding, cosmic-ray removal, and bad pixel correction was performed with {\tt IRAF} and {\tt MIDAS} tools. As the images were mainly obtained in a grey time, they were partially contaminated by a nonuniform background caused by  a small OSIRIS detector filter wheels inclination.\footnote{http://www.gtc.iac.es/instruments/osiris/media/OSIRIS-USER-MANUAL\_v2.1.pdf} In order to eliminate the contamination, we performed illumination corrections for each observational set. Finally, using a set of unsaturated stars, we aligned all the individual exposures in each band to the best ones obtained at the highest quality seeing conditions. The resulting combined images have mean seeings of  0\farcs93 and 0\farcs96, mean airmasses of 1.62 and 1.77, and total integration times of 7.55 and 18.09 ks, for $g'$ and $i'$ bands, respectively.

For photometric calibrations, we used photometric standards SA 112 805 and PG 1047$+$003a for the $g'$ band, and SA 104 428 for the $i'$ band, which were observed the same nights as our target. The atmospheric extinction  coefficients $k_{g'}= 0.15\pm 0.02$ and $k_{i'}= 0.04\pm 0.01$ were taken from  2014 OSIRIS/GTC broad-band imaging calibration notes.\footnote{http://www.gtc.iac.es/instruments/osiris/media/CUPS\_BBpaper.pdf} The derived magnitude zero-points for $g'$ and $i'$ images in CCD2, where our target was exposed, are 28\fm79$\pm$0\fm04 and 28\fm71$\pm$0\fm04, respectively. The uncertainties include the statistical measurement errors, $\approx4\%$ stellar flux variations from exposure to exposure of the target field, the atmospheric extinction  coefficient uncertainties, and, in the case of the $g'$-band, the zero-point variations from night to night. The $g'$-band zero-point is fully consistent with the mean values for photometric/clear nights presented in the OSIRIS/GTC calibration notes, while the $i'$ zero-point is marginally, by 1$\sigma$, lower than the respective mean values, which may indicate that the $i'$ observations  were slightly affected by dust in the atmosphere as typical for the GTC site during Summer time. Here, we ignore colour terms in the instrumental to AB magnitude transformations. Their contributions were evaluated to be always within the estimated error budgets and did not affect our results.

\begin{table}[t]
  \caption{Log of the GTC/OSIRIS observations of the Calvera field.}\label{t:log}
  \begin{center}
    \begin{tabular}{lllll}
      \hline\hline
      Date           &  Band         & Exposure               &  Airmass      & Seeing          \\
                     &               & [s]                    &               & [arcsec]        \\
      \hline
      2014-04-04     &  $g'$         & 4~$\times$~686         & 1.59$-$1.62   &  0.9$-$1.1      \\
      2014-06-04     &  $i'$         & 35~$\times$~134        & 1.64$-$1.79   &  0.7$-$1.0      \\
      2014-06-26     &  $g'$         & 7~$\times$~686         & 1.58$-$1.63   &  0.7$-$0.9      \\
      \hline
    \end{tabular}
  \end{center}
\end{table}

\subsection {Gemini-North archival data} \label{sec:gemini-data}

In our analysis, we also used the archival Gemini-North Multiobject Spectrograph (GMOS-N) observations obtained in 2006 December in the $g$ band with the exposure of $4\times600$ s \citep{rutledge2008ApJ}. Four dithered exposures  were retrieved from the Gemini archive and reduced using the {\tt IRAF GMOS} pipeline and official calibration products (bias, flats, bad pixel map, etc.). The image scale and unvignetted field size are 0\farcs145 pixel$^{-1}$ and $5\farcm4\times 5\farcm4$, respectively. Calvera was exposed on the central of the three GMOS CCDs. The mean seeing and airmass on the combined image are 1\farcs3 and 2.3, respectively.

\subsection{\textit{Chandra} and \textit{XMM-Newton} archival data}\label{sec:x-ray-data}

\begin{table*}[t]
  \caption{X-ray spectral observations of Calvera.}\label{t:data}
  \begin{center}
    \begin{tabular}{ccccccccc}
      \hline\hline
      DateObs    & Exp.   & Inst.     & Mission             & ObsID      & ObsMode & Observer   & Phot.$^{a}$ \\

      [UT]       & [s]    &           &                     &            &         &            &             \\
      \hline
      2009-08-31 & 11351  & EPIC-pn   & \textit{XMM-Newton} & 0601180101 & SW      & S. Zane    & 5958        \\
      2009-10-10 & 19477  & EPIC-pn   & \textit{XMM-Newton} & 0601180201 & SW      & S. Zane    & 9846        \\
      2009-08-31 & 6273   & EPIC-MOS1 & \textit{XMM-Newton} & 0601180101 & FW      & S. Zane    & 694         \\
      2009-10-10 & 22469  & EPIC-MOS1 & \textit{XMM-Newton} & 0601180201 & FW      & S. Zane    & 2481        \\
      2013-02-18 & 17093  & ACIS-S3   & \textit{Chandra}    & 15613      & CC      & J. Halpern & 2429        \\
      2013-02-12 & 19679  & ACIS-S3   & \textit{Chandra}    & 13788      & CC      & J. Halpern & 2675        \\
      2008-04-08 & 26430  & ACIS-S3   & \textit{Chandra}    & 9141       & VFAINT  & D. Fox     & 4609        \\
      \hline
    \end{tabular}
  \end{center}
  \tablenotetext{}{$^a$ Number of source counts in the 0.3--7.0 keV range.}
\end{table*}

For the X-ray spectral analysis, we used archival \textit{Chandra} Advanced CCD Imaging Spectrometer (ACIS) and \textit{XMM-Newton} European Photon Imaging Camera (EPIC) data (Table~\ref{t:data}). The \textit{XMM-Newton} data were processed with the {\tt XMM-SAS v.13.5.0} software. We selected single and double pixel events (PATTERN~$\leq$~4) for the EPIC-pn and single to quadruple-pixel events (PATTERN~$\leq$~12) for the EPIC-MOS1 data. We did not use the MOS2 data for spectral analysis since MOS2 was in the timing mode which is not accurately calibrated for spectral analysis. We removed periods of background flares using 10--12 keV and 12--14 keV light curves for the MOS1 and pn data, respectively. Using {\tt evselect} tool, we extracted spectra from the 30\asec~aperture centered at the Calvera position which encapsulates most of the Calvera emission. We used {\tt SAS} tasks {\tt rmfgen} and {\tt arfgen} to generate  redistribution matrix and ancillary response files, respectively. The resulting exposure times and number of source counts (in the 0.3--7.0 keV range) after background subtraction are listed in Table~\ref{t:data}. The total number of the source counts obtained with \textit{XMM-Newton} in the 0.3--7.0 keV range is 18979.

In the \textit{Chandra} spectral observation, Calvera was exposed on the ACIS-S3 CCD chip. We reprocessed these data using the {\tt CIAO v.4.7 chandra\_repro} tool with {\tt CALDB v.4.6.9}. For the data obtained in the VFAINT mode (ObsID 9141), we extracted the pulsar photons from 1\farcs5 aperture centered at the Calvera position with the {\tt CIAO specextract} tool. For the data obtained in the CC mode (ObsIDs 13788 and 15613), we extracted spectrum from the five columns near the pulsar position. The total number of the source counts obtained with \textit{Chandra} in the 0.3--7.0 keV range is 9713 (see Table~\ref{t:data}). For the \textit{XMM-Newton} and \textit{Chandra} data, background was taken from regions free from any sources. All spectra were grouped to ensure at least 25 counts in each energy bin. For the X-ray image analysis, we used the archival \textit{Chandra} High Resolution Camera data obtained with the micro-channel plate imaging detector I (HRC-I)\footnote{ObsId 15806, exposure $\approx 30$ ks, PI J. Halpern.} in 2014 April almost simultaneously with the GTC data \citep{Halpern2015}. The HRC-I data provides subarcsecond resolution useful for optical counterpart identification. The data were reprocessed with the {\tt chandra\_repro} tool and the exposure map corrected image was then created.

\section{Optical and X-ray astrometry}\label{sec:astrometry}

Precise astrometric referencing between the \textit{Chandra}/HRC-I and GTC images is crucial to search for the Calvera optical counterpart. The only appropriate reference source for that is USNO-B1.0 1693$-$0051234 which X-ray counterpart CXOU J141259.4+791958 is located $2\amin$ southward of Calvera \citep{rutledge2008ApJ}. Unfortunately, it is strongly saturated in GTC images. The same is true for other bright astrometric standards in the field. There are also some geometric distortions in GTC images toward CCD boundaries outside a few arcminutes from the Calvera position. Therefore, we used the Gemini image, which has negligible geometric distortions and where the reference source and most other standards are unsaturated, as a proxy to obtain the astrometric solution. Using twenty unsaturated stars around Calvera from the USNO-B1 astrometric catalogue and the {\tt IRAF ccmap} task, we referenced the Gemini image to the World Coordinate System (WCS) with formal {\sl rms} uncertainties of the astrometric fit $\Delta$RA~$\approx 0\farcs11$ and $\Delta$Dec~$\approx 0\farcs13$. This is consistent with the nominal catalogue uncertainty of $\approx 0\farcs2$ and results in a conservative 1$\sigma$ WCS referencing uncertainty of 0\farcs23 for the whole Gemini image. We then performed referencing of the central part of the GTC $g'$ image containing Calvera to the Gemini image. We selected nine point-like unsaturated secondary astrometric standards well detected in both images within $\sim 2\amin$ around the Calvera position. The {\sl rms} reference uncertainties of 0\farcs092 and 0\farcs044 for RA and Dec, respectively, yielded a  1$\sigma$ WCS referencing uncertainty of $\approx 0\farcs24$ for this region. The GTC $i'$ image was then tied to the $g'$ image with the uncertainties of $\Delta \rm RA = 0\farcs042$ and $\Delta \rm Dec = 0\farcs055$.

The referencing of the HRC-I image to the Gemini one is not straightforward, since only one reference source is available. Fortunately, the source USNO-B1.0 1693$-$0051234 is extragalactic and its proper motion is negligible \citep{Halpern2015}. Comparing the WCS-referenced Gemini image and the HRC-I image with nominal \textit{Chandra} WCS referencing, we found a significant offset of 0\farcs396 between the optical and X-ray coordinates of the reference source, which is consistent with what was reported by \citet{Halpern2015}. The offset is within the nominal 90\% astrometric uncertainty of the HRC-I image of 0\farcs6, however fine tuning of Gemini-HRC-I referencing is possible. Following \citet{Halpern2015}, we shifted the HRC-I image eliminating the offset. The centroid positions of the reference source in the HRC-I and Gemini images have uncertainties of 0\farcs025 and 0\farcs005, respectively, for both coordinates, therefore the image shift is determined with the accuracy of 0\farcs025.

In the HRC-I observations, the reference source was placed on the optical axis to ensure the most precise referencing \citep{Halpern2015} and therefore Calvera was observed 2\arcmin\ off-axis. For the off-axis sources, there are additional reference errors related to the plate rotation and scale, which can not be accounted for with only one reference source. The \textit{Chandra} roll angle uncertainty is $\approx 25\asec$ and hence it results in a systematic error of 0\farcs015 for a source 2\amin\ off-axis.\footnote{http://cxc.harvard.edu/cal/ASPECT/roll\_accuracy.html} We examined possible non-linearity of the HRC plate scale using some archival data with sufficient number of reference sources and found that it introduces $\lesssim 0\farcs05$ astrometric error within 2\amin\ from the optical axis. Details of the analysis are given in the Appendix. We took these values into account in resulting uncertainty of the HRC-I referencing. Combining the referencing uncertainties with the HRC-I Calvera centroid position uncertainty of 0\farcs04, we found that the Calvera position can be localized on the Gemini image with the 1$\sigma$ uncertainty of 0\farcs07.

Accounting for all the uncertainties, the 1$\sigma$ accuracy of referencing the HRC-I image within 2\amin\ from the optical axis to WCS is 0\farcs24 for both coordinates. The same accuracy holds for WCS referencing of the GTC $g'$- and $i'$-image fragments containing 2\amin\ vicinity of Calvera. The resulting X-ray position  of Calvera is shown in Table~\ref{t:phot}. The reference source USNO-B1.0 1693$-$0051234 position in our optical images is RA~= 14:12:59.42(8)\footnote{Here and below all uncertainties given in brackets refer to the last significant digits quoted.} and Dec~= +79:19:58.80(24). These positions are consistent within uncertainties with those reported by \citet{Halpern2015} who used similar approach but with the MDM Observatory 2.4-m Hiltner telescope $R$-band image of the field, providing independent confirmation of their results.

\begin{figure}[th]
  \setlength{\unitlength}{1mm}
  \begin{picture}(105,115)(0,0)
    \put (6,56)   {\includegraphics[scale=0.55]{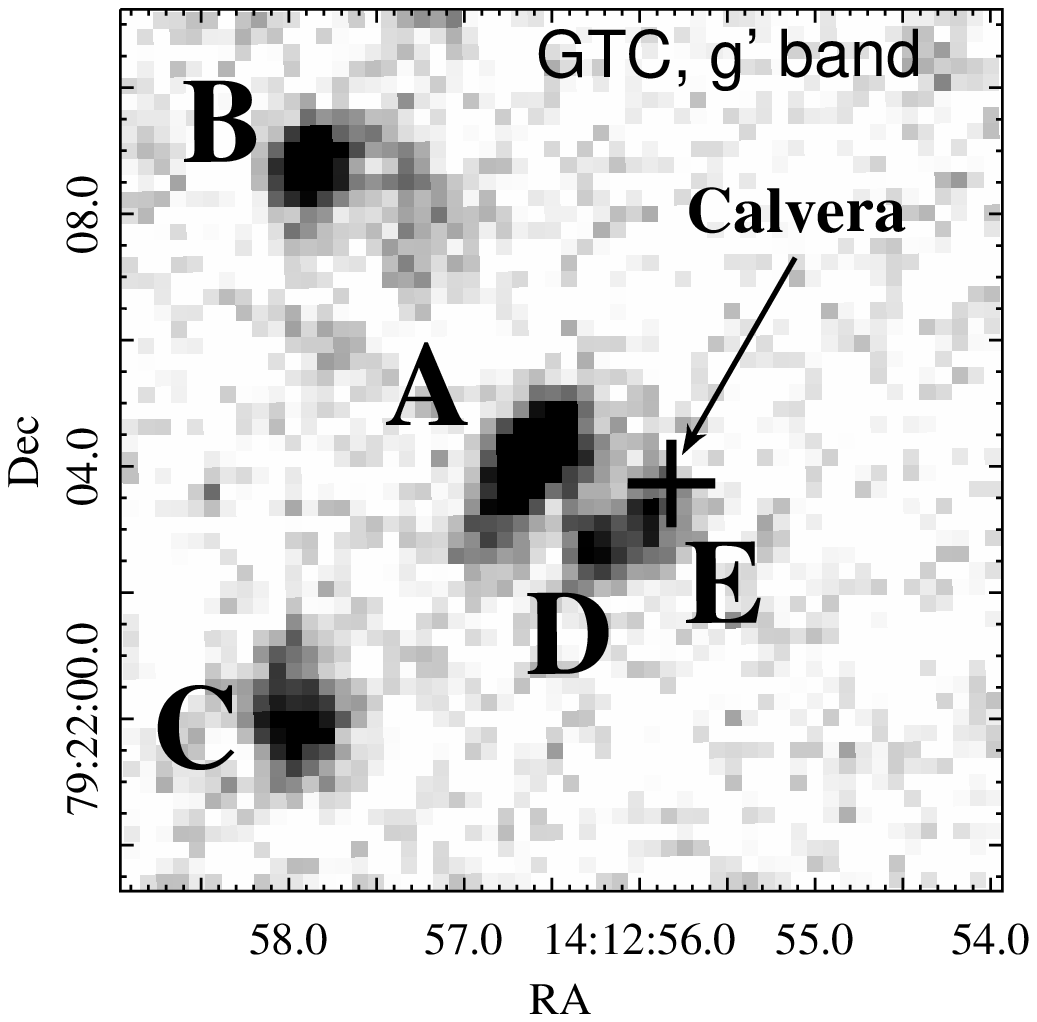}}
    \put (90,56)  {\includegraphics[scale=0.55]{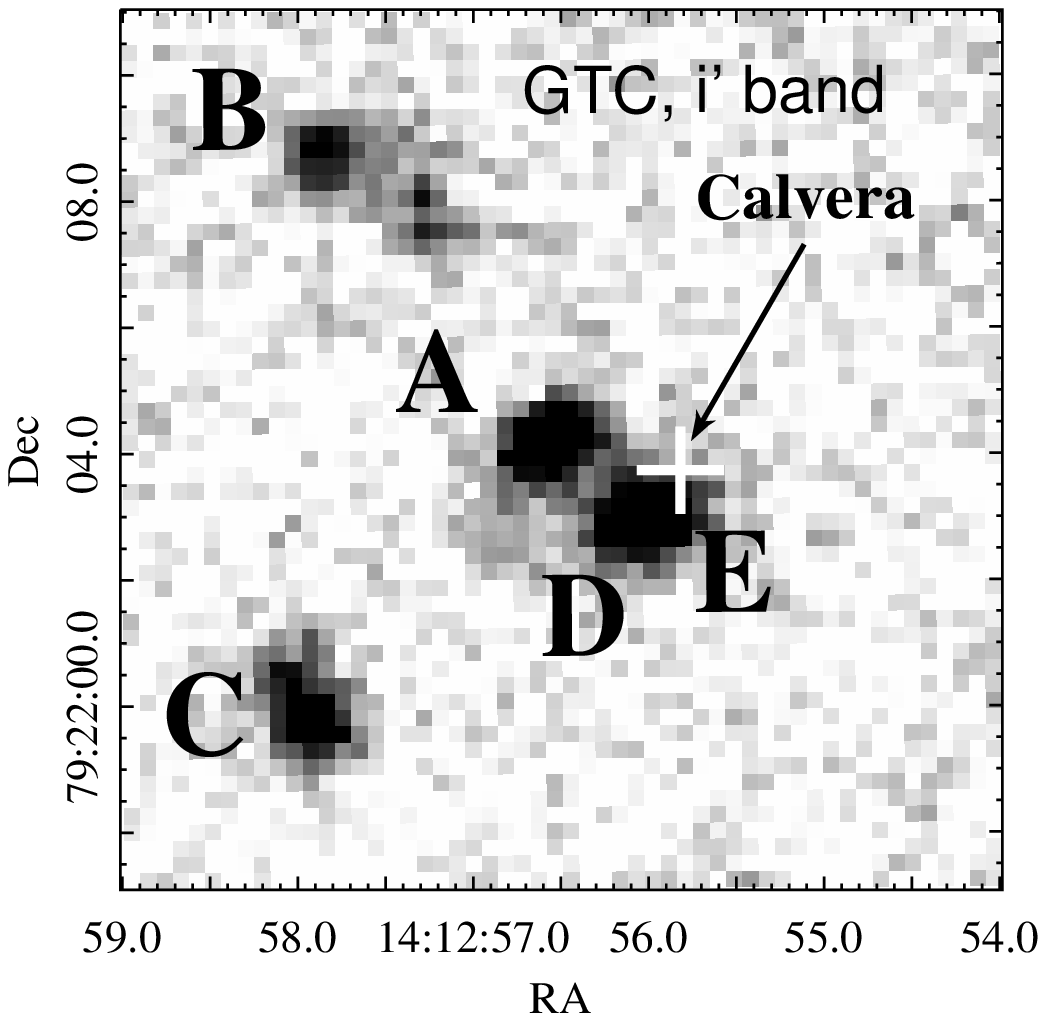}}
    \put (5.,-3)  {\includegraphics[scale=0.55]{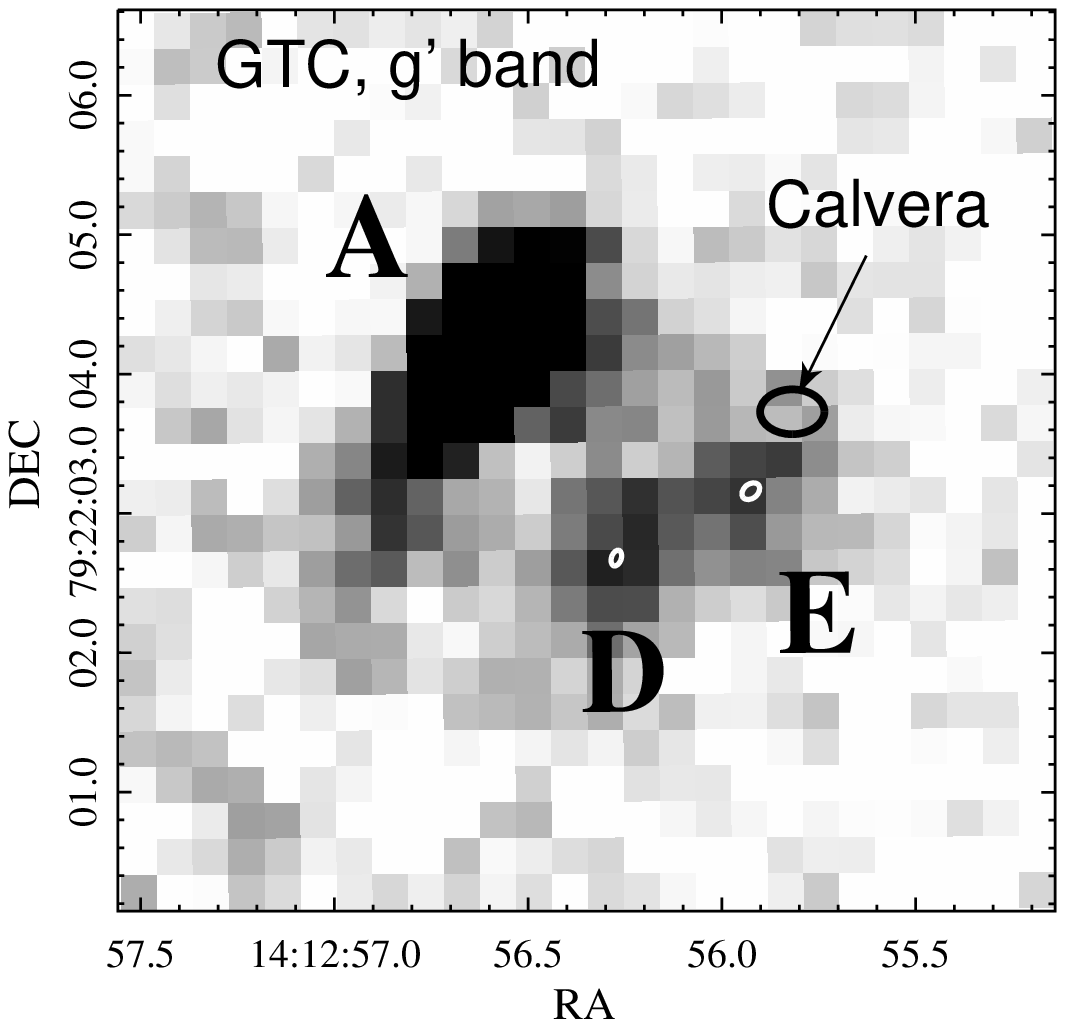}}
    \put (90.,-3) {\includegraphics[scale=0.55]{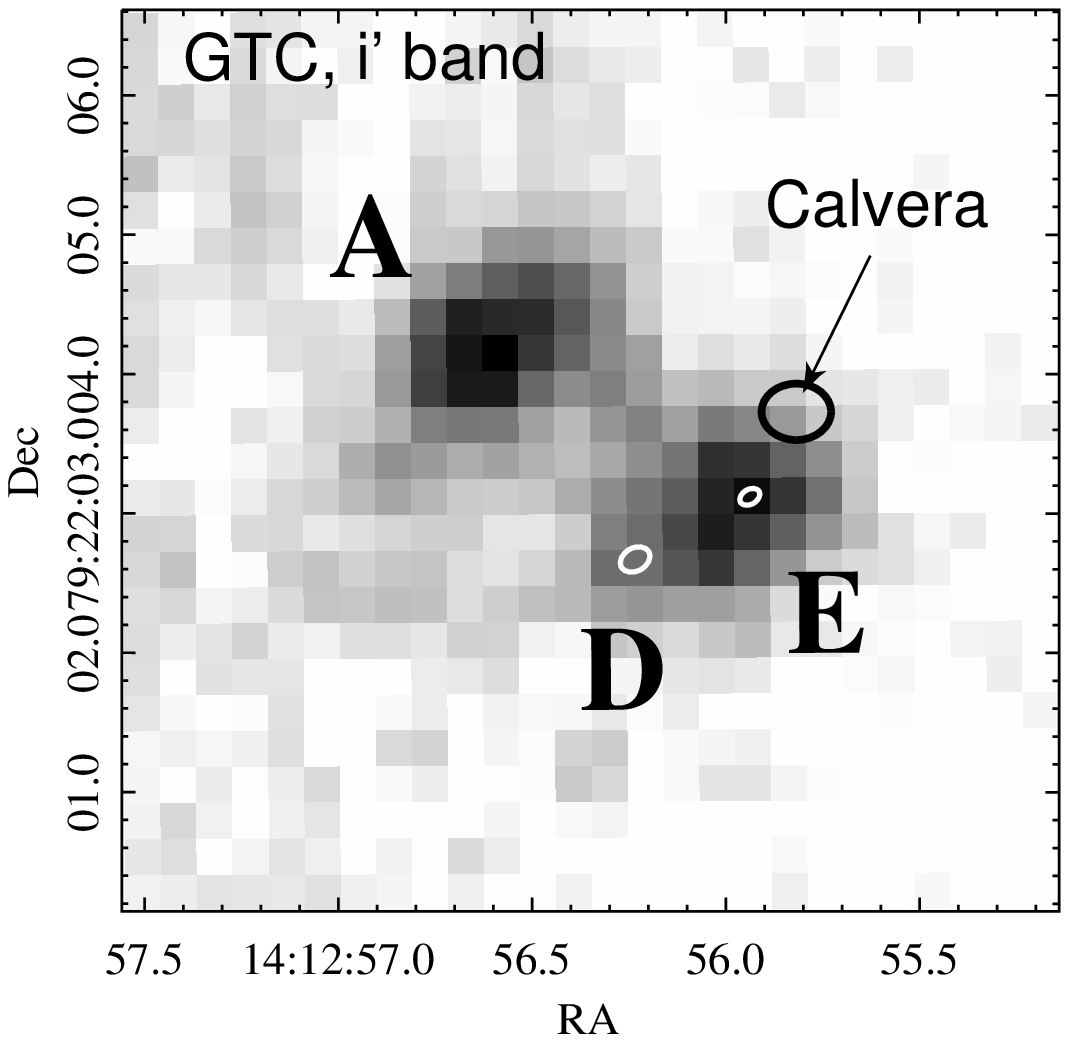}}
  \end{picture}
  \caption{{\sl Top:} $14\asec\times14\asec$ images of the Calvera field obtained in the optical $g'$ and $i'$ bands with the GTC. Cross shows position of Calvera. Labels $A$, $B$, $C$, $D$, and $E$ mark optical sources in the field. {\sl Bottom:} Smaller fragments of the same images zooming on a vicinity of the Calvera X-ray position. Ellipses show 90\% position uncertainties for sources $D$, $E$, and Calvera.}\label{fig:optyu}
\end{figure}

\section{Searching for the Calvera optical counterpart}\label{sec:search}

GTC image fragments of the Calvera field are shown in the top panels of Figure~\ref{fig:optyu}. The Calvera X-ray position is marked by the cross. The nearby background optical objects $A$, $B$, and $C$ are notated following \citet{rutledge2008ApJ}. $D$ and $E$ mark new sources detected with the GTC with the signal-to-noise ratio $S/N > 10$. $D$ and $E$ are well resolved in the $g'$ band while they are blended in the $i'$ image. Positions of all these sources are shown in Table~\ref{t:phot}. As seen, object $E$ is the closest source to Calvera which projects onto the north-west wing of the object $E$ spatial profile. Therefore precise measurement of the offset between them is important to verify whether $E$ can be associated with the pulsar.

The immediate vicinity of Calvera, which includes sources $A$, $D$, and $E$, is  enlarged in the bottom panels of Figure~\ref{fig:optyu}. The positions of sources $E$ and $D$ were measured with the point spread function (PSF) fit in each of GTC images, using {\tt IRAF/daophot/nstar} task. The PSFs were generated using the {\tt IRAF/daophot/psf} task for eleven-pixel radius, where the bright isolated unsaturated stars selected for the PSF construction merge with the background. An optimal PSF fit radius was chosen to be about three pixels, in accordance with seeing conditions.

To estimate the position uncertainties, we performed Monte Carlo simulations of synthetic data sets. That is, using the PSF model we simulated synthetic images of sources $A$, $E$, and $D$ assuming Poisson distribution of total brightness of each source. The background contribution in each pixel was drawn from the Poisson distribution basing on the respective pixel value in the initial {\tt IRAF/daophot/nstar} fit residual image. In this way, we account for the possible background variations over the image. Then, the uncertainties were derived from the distribution of sources positions measured on the synthetic images. The resulting $1\sigma$ uncertainties are $\la 0\farcs03$ and $\la 0\farcs05$ for the $g'$ and $i'$ images, respectively, and are consistent with estimates given by the relation from \citet{bobroff1986RScI} for specific seeing and $S/N$ values. The 90\% uncertainties of $E$ and $D$ positions are shown by white ellipses in the bottom panels of Figure~\ref{fig:optyu}. The positions of $D$ and $E$ in both bands are consistent within statistical and $i'-g'$ referencing uncertainties.

Black ellipses in the bottom panels of Figure~\ref{fig:optyu} show the 90\% uncertainties of the Calvera HRC-I position as referenced to the GTC images. The position uncertainties are combined from the 0\farcs07 uncertainty of the HRC-I Calvera position as referenced to the Gemini image, the Gemini$-$GTC referencing uncertainty ($\Delta {\rm RA}=0\farcs092$ and $\Delta { \rm Dec} = 0 \farcs 044$) and, for the $i'$ image, $i'-g'$ transformation accuracy ($\Delta {\rm RA}=0\farcs042$ and $\Delta {\rm Dec} = 0 \farcs 055$). The resulting sizes of the ellipses are $0\farcs24\times 0\farcs17$ and $0\farcs26\times 0\farcs19$ for $g'$- and $i'$-band images, respectively. As seen, source $E$ can hardly be an optical counterpart of the Calvera pulsar. Its offset from the Calvera position is $0\farcs67 \pm 0\farcs15$ which corresponds to $\approx 4.5\sigma$ significance or $5 \times 10^{-5}$ probability of the coincidence according to the Rayleigh distribution.

\section{Photometry of the GTC images}\label{sec:phot-anal}

\begin{table}[t]
  \caption{Measured magnitudes and positions.}\label{t:phot}
  \begin{center}
    \begin{tabular}{lllll}
      \hline\hline
      Source     & RA (J2000)     & Dec (J2000)       &  $g'$            & $i'$                       \\
                 &                &                   &  [mag]           & [mag]          \\
      \hline
      $A$        & 14:12:56.60(8) & +79:22:04.17(24)  & 24.71(4)         & 24.20(11)        \\
      $B$        & 14:12:57.86(8) & +79:22:08.77(24)  & 25.17(3)         & 24.53(5)       \\
      $C$        & 14:12:57.97(8) & +79:21:59.98(24)  & 25.47(6)         & 24.30(5)       \\
      $D$        & 14:12:56.27(8) & +79:22:02.68(24)  & 26.10(5)         & 25.37(13)        \\
      $E$        & 14:12:55.93(8) & +79:22:03.16(24)  & 26.22(6)         & 24.17(11)        \\
      Calvera    & 14:12:55.82(8) & +79:22:03.73(24)  & $\ga 27.87$      & $\ga 26.84$      \\
      \hline
    \end{tabular}
    \tablenotetext{}{{\bf Notes.} Measured magnitudes and positions of point-like optical sources detected in the Calvera vicinity and labeled in Figure~\ref{fig:optyu} and the Calvera $3\sigma$ brightness upper limits and X-ray position. Numbers in brackets are 1$\sigma$ uncertainties referring to the last significant digits quoted.}
  \end{center}
\end{table}

Since no reliable optical counterpart of Calvera was found, only an upper limit on its optical flux can be placed. In the source-free regions near the Calvera position in the GTC images, the 3$\sigma$ point-source detection limits are  $g'>27.93$ and $i'>26.84$. However, Calvera locates in the wings of the source $E$, which is also  blended  with the source $D$. Therefore, PSF photometry is needed to robustly estimate Calvera brightness upper limits. We performed PSF photometry using the {\tt IRAF/daophot allstar} task with the PSF fit radius of three pixels. Annulus and dannulus of eleven and ten pixels were typically used to extract local backgrounds. Using aperture photometry for a large set of field stars, we checked that in our data the {\tt allstar} task does not result in a biased PSF photometry. Aperture corrections were estimated based on the photometry of bright unsaturated field stars and photometric standards using aperture radii up to 20\asec. For the $g'$ band, the correction was $0\fm45\pm0\fm01$, while for the $i'$ band with a worse seeing, it was 0\fm60$\pm$0\fm01. The magnitude errors were estimated accounting for the statistical measurement errors, a magnitude dispersion in the star iterative subtraction process within the {\tt allstar} task, calibration zero-points, atmospheric coefficient, and aperture correction uncertainties. The PSF subtraction of the stellar-like sources $D$ and $E$ is perfect allowing for the reliable Calvera upper limit estimate. Calvera brightness 3$\sigma$ upper limits estimated using background standard deviations on its positions in the star-subtracted images are $g'> 27.87$ and $i'>26.89$. Within the zero-point and aperture correction uncertainties, these values are compatible with the upper limits given above for a wider region free of any optical source. For  conservative resulting estimates, we accepted the brightest of the limits obtained in each band, i.e., $g'>27.87$, $i'>26.84$. The photometry results for the sources $A-E$ are summarized in Table~\ref{t:phot}. In the $g'$ band, $D$ and $E$ have similar magnitudes, while $E$ becomes significantly brighter than $D$ in the $i'$ band. For the source $A$, our result in the $g'$ band is consistent within uncertainties with the estimates by \citet{rutledge2008ApJ} based on the calibration using ten USNO stars in the field. On the other hand, the brightnesses of fainter $B$ and $C$ sources were significantly underestimated in their work by a value of $\sim$0\fm4, probably due to incorrect aperture correction. According to Table~\ref{t:phot}, object $E$ is a magnitude redder than other sources. This serves as an additional argument against $E$ being the Calvera optical counterpart since most pulsars are blue objects in the optical band.

\section{X-ray spectral analysis}\label{sec:spect}

Below we consider the phase-averaged X-ray spectra only. Previous analyses of the X-ray data showed that any one-component model (absorbed) cannot satisfactory describe the spectrum of Calvera \citep{zane2011MNRAS,halpern2013ApJ}. Models with a non-thermal component are excluded since they require absorption several times the total Galactic absorption in the direction to Calvera to fit the soft part of the spectrum in \textit{XMM-Newton} data \citep{zane2011MNRAS}. Thus, the soft part of the Calvera spectrum is supposed to be thermal. \citet{halpern2013ApJ} preferred a model of two hot spots with blackbody spectra which fit the phase-averaged spectrum and can explain dependence of the pulsed fraction and pulse shape on photon energy. \citet{zane2011MNRAS} showed that the phase-averaged spectrum can be also described by the sum of two hydrogen atmosphere models NSA \citep{pavlov1995lns,zavlin1996AsAp}.

\citet{zane2011MNRAS} examined the phase-averaged spectra for the presence of an absorption feature. They found that the fit improvement when multiplying the two-component thermal model by an absorption edge is statistically significant. In addition, they showed that the fit with the NSA model with the absorption edge is also acceptable. The presumed presence of an emission line at $\sim 0.5$ keV was pointed out earlier by \citet{shevchuk2009ApJ} based on analysis of the \textit{Chandra}/ASIS-S data, but the line significance was inconclusive.

\begin{table*}[t]
  \caption{Best-fit parameters for NSMAX models.}\label{t:x-fit}
  \begin{center}
    \begin{tabular}{l*{8}{c}}
      \hline\hline
      Model    & $N_{\rm H}$            & $T^{\infty}$             &  $R/D$                 &     $E_{0}$             &   FWHM                   &   EW                  & $\chi^{2}$/d.o.f.  \\
         & [10$^{20}$ cm$^{-2}$]  & [10$^{6}$ K]             &  [km/kpc]              &     [keV]               &   [keV]                  &     [keV]             &                    \\
      \hline
      \multicolumn{7}{c}{}\\
      \multicolumn{8}{c}{without absorption line}\\
      \multicolumn{7}{c}{}\\
      123100   & $<1.1$                 & $0.84^{+0.03}_{-0.02}$   & $5.9^{+0.6}_{-0.6}$    & $-$                     & $-$                      & $-$                   & 943/895            \\ % ok
      \multicolumn{7}{c}{}\\
      123190   & $<0.3$                 & $1.34^{+0.01}_{-0.02}$   & $2.4^{+0.2}_{-0.1}$    & $-$                     & $-$                      & $-$                   & 977/895            \\ % ok
      \multicolumn{7}{c}{}\\
      1200     & $<0.4$                 & $1.25^{+0.02}_{-0.01}$   & $2.4^{+0.1}_{-0.2}$    & $-$                     & $-$                      & $-$                   & 962/895            \\ % ok
      \multicolumn{7}{c}{}\\
      \multicolumn{8}{c}{with absorption line}\\
      \multicolumn{7}{c}{}\\
      123100   & $2.0^{+0.9}_{-0.9}$    & $0.79^{+0.03}_{-0.03}$   & $7.4^{+1.1}_{-1.1}$    & $0.74^{+0.03}_{-0.03}$  &  $0.20^{+0.10}_{-0.08}$  & $0.03^{+0.02}_{-0.01}$ & 906/892       \\ % ok
      \multicolumn{7}{c}{}\\
      123190   & $< 1.3$                & $1.29^{+0.03}_{-0.05}$   & $2.7^{+0.4}_{-0.2}$    & $0.74^{+0.02}_{-0.02}$  &  $0.25^{+0.07}_{-0.07}$  & $0.05^{+0.02}_{-0.02}$ & 907/892       \\ % ok
      \multicolumn{7}{c}{}\\
      1200     & $1.2^{+1.2}_{-0.9}$    & $1.18^{+0.05}_{-0.05}$   & $2.9^{+0.4}_{-0.4}$    & $0.73^{+0.03}_{-0.03}$  &  $0.24^{+0.10}_{-0.07}$ & $0.05^{+0.02}_{-0.02}$  & 902/892       \\ % ok
      \multicolumn{7}{c}{}\\
      \hline
    \end{tabular}
    \tablenotetext{}{{\bf Notes.} Temperatures $T^{\infty}$ are given as measured by a distant observer. Ratios $R/D$ are weighted averages of corresponding values for various instruments and their uncertainties therefore account for both statistical and systematic errors. EW is equivalent width. All errors correspond to 90\% HPD credible intervals derived via MCMC.}
  \end{center}
\end{table*}

We considered various atmosphere models and found that spectra can be described by one-component magnetized hydrogen atmosphere model NSMAX \citep{ho2008ApJS} which has not yet been applied to the Calvera spectra. We therefore focus on this particular model in our analysis. It has certain advantages over the older NSA model. In particular, the NSMAX model accounts for the effects of partial ionization in the opacities of the atmosphere. Moreover, there are public available NSMAX models which incorporate smooth (dipole) inhomogeneities of the surface temperature and the magnetic field over the NS surface.

\begin{figure}[t]
  \setlength{\unitlength}{1mm}
  \begin{picture}(160,60)(0,0)
    \put (-3.0,0.0) {\includegraphics[scale=0.45]{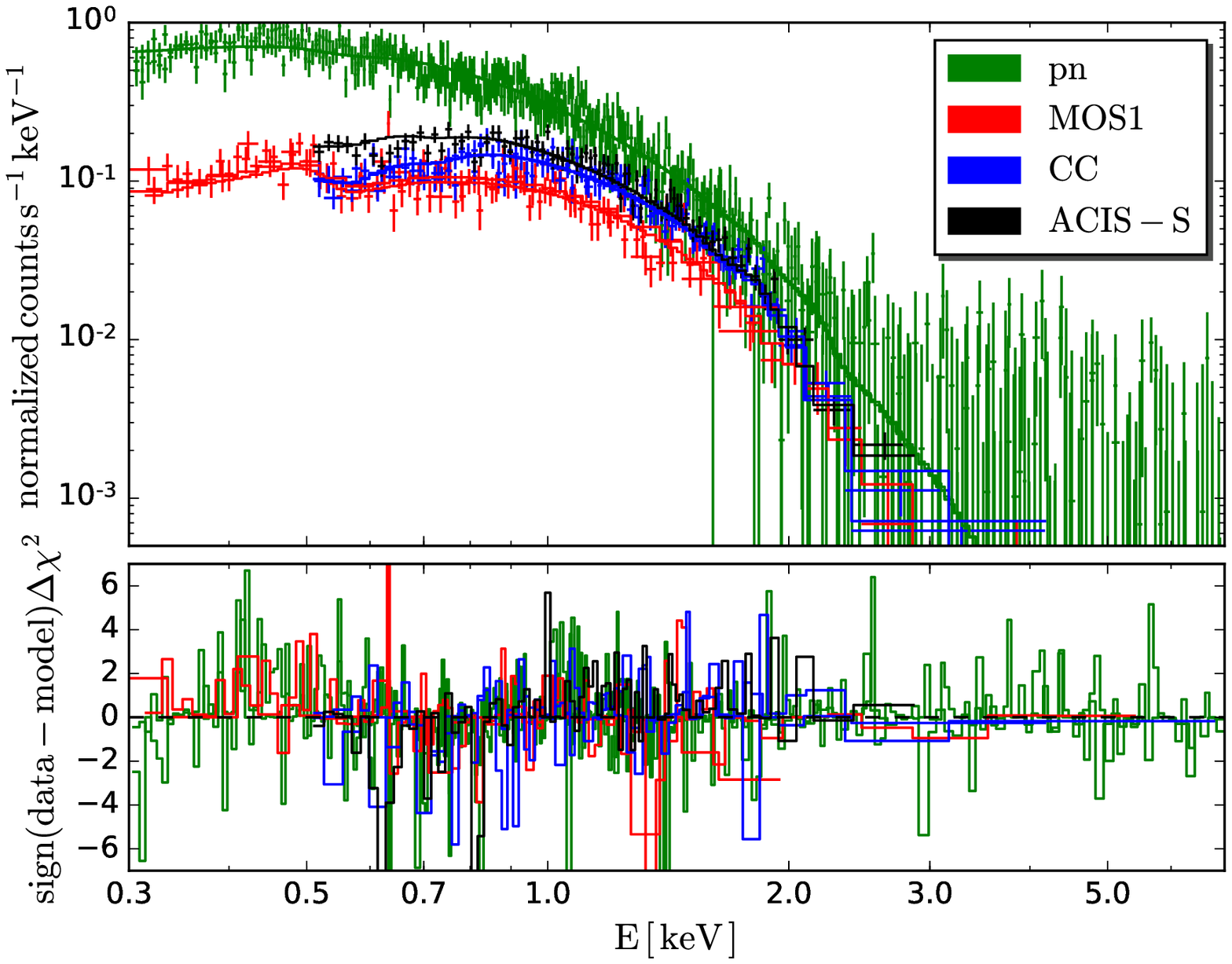}}
    \put (80.0,0.0) {\includegraphics[scale=0.45]{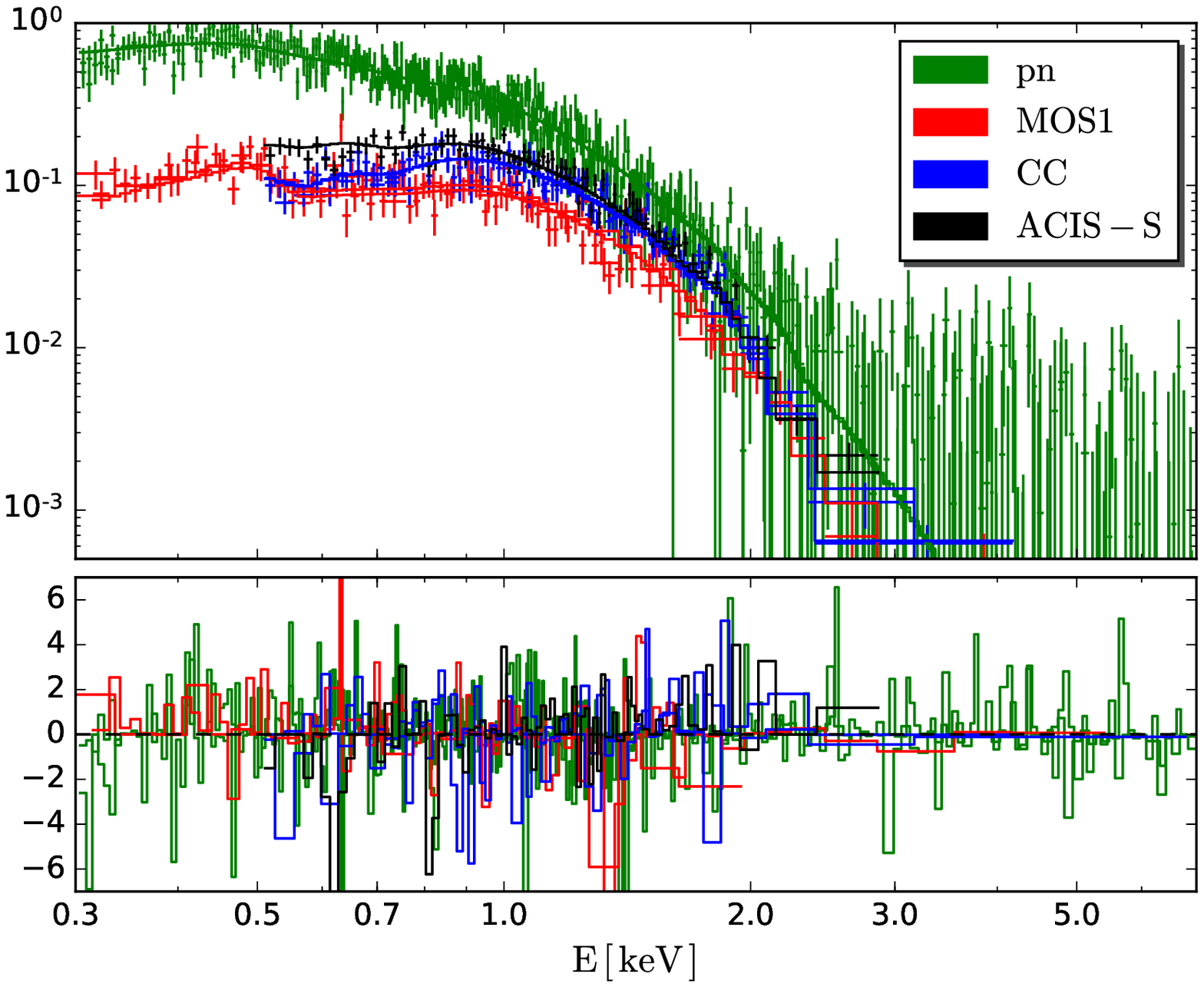}}
  \end{picture}
  \caption{The X-ray spectra of Calvera and the best-fit hydrogen atmosphere model with uniform temperature (NSMAX 1200) without ({\sl left}) and with ({\sl right}) absorption feature.}
  \label{fig:x-ray_nsmax}
\end{figure}

The \textit{XMM-Newton} data in the 0.3$-$7~keV range and the \textit{Chandra} data in the 0.5$-$7~keV range were fitted simultaneously. For the spectral fitting, we used the {\tt Xspec v.12.9.0} spectral fitting package \citep{arnaud1996ASPC} connected to the {\tt Python} Markov Chain Monte Carlo (MCMC) package {\tt emcee} \citep{foreman-mackey2013PASP} through the {\tt PyXspec Python} front-end to {\tt Xspec}. This approach allowed us to effectively sample the posterior distribution and, in particular, to obtain the fit quality and  uncertainties of parameters.

In upper rows of Table~\ref{t:x-fit}, we show the best-fit parameters for the three models from the NSMAX series, 1200, 123190, and 123100. The parameters are the hydrogen column density $N_{\rm H}$, the NS radius to distance ratio $R/D$, and effective temperature as seen by a distant observer $T^{\infty}$. The models also depend on the gravitational redshift at the NS surface but since it was not actually constrained from the fit, we just marginalized posterior distributions over it. The interstellar absorption is accounted for by the TBABS model with abundances from \citet{wilms2000ApJ}. The NSMAX  1200 model assumes the constant radial magnetic field of 10$^{12}$~G, while the models 123100 and 123190 account in a particular way for the dipole variations of the magnetic field ($B=10^{12}$~G at the magnetic equator) and the temperature over the surface \citep{ho2008ApJS}. For the NSMAX 123100 model, the NS is assumed to be viewed from the magnetic pole, while for the NSMAX 123190 model the star is viewed in the orthogonal direction. In order to account for the cross-calibration uncertainties between the \textit{XMM-Newton} pn, MOS1, and \textit{Chandra} detectors,\footnote{https://heasarc.gsfc.nasa.gov/docs/heasarc/caldb/caldb\_xcal.html} we let $R/D$ vary independently. The parameter uncertainties in Table~\ref{t:x-fit} are 90\% highest posterior density (HPD) credible intervals derived from the marginalized posterior distributions of respective parameters \citep[e.g.,][]{protassov2002ApJ, GelmanBook}. For $R/D$, we give the weighted average value with uncertainties accounting for the statistical and cross-calibration systematic errors.

\begin{figure}[t]
  \begin{center}
    \includegraphics[scale=0.36,clip]{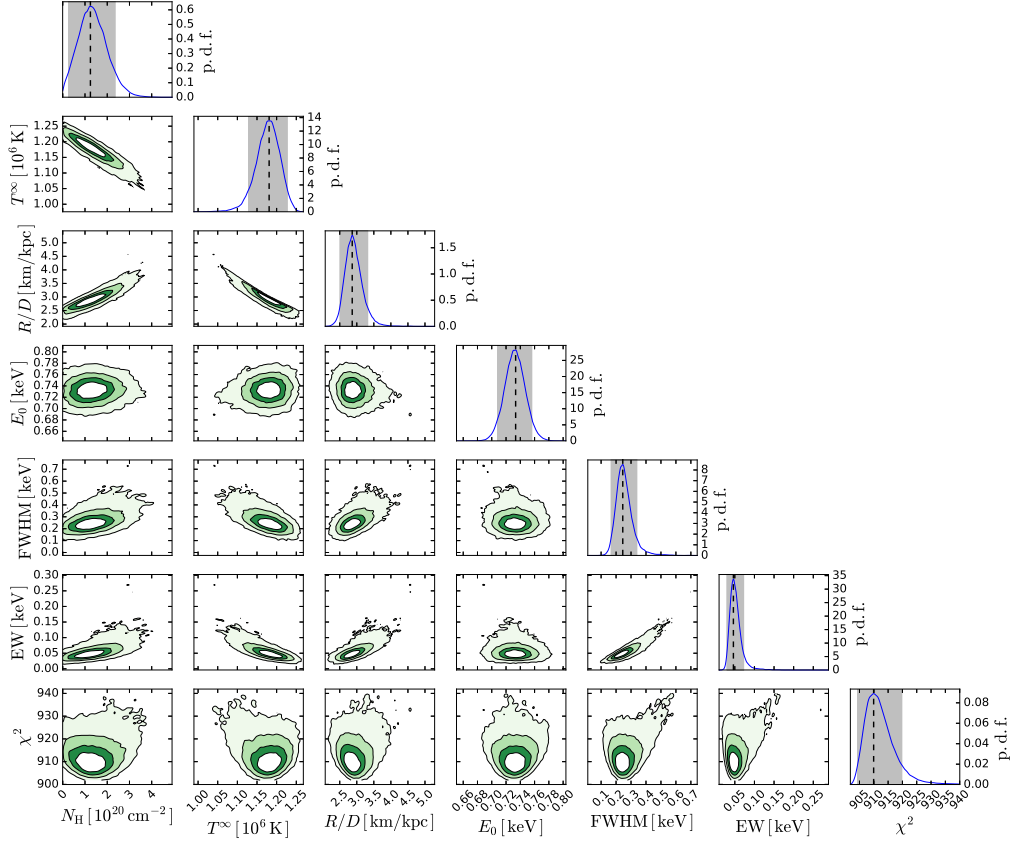}
  \end{center}
  \caption{1D and 2D marginalized posterior distributions for the NSMAX(1200)$\times$GABS model. Gray-filled regions correspond to 90\% HPD credible intervals. Contours are for 40\%, 68\%, 90\%, and 99\% levels.}
  \label{fig:nsmaxgabs_triangle}
\end{figure}

In addition to traditional inspection of the $\chi^2$ value given in the last column in Table~\ref{t:x-fit}, we assessed the goodness-of-fit test by means of the posterior predictive check \citep{GelmanBook}. We found that each of the models in Table~\ref{t:x-fit} fit the data equally well. The example of the spectrum and best-fit residuals is shown in the left panel of Figure~\ref{fig:x-ray_nsmax} for the NSMAX 1200 model.

It is seen, that a singe-component NSMAX 1200 fit is good, but there is the wavy pattern seen in the fit residuals between 0.4 and 1.0 keV with the dip at about 0.7 keV which probably points to the presence of a spectral feature. The same wavy pattern is seen for other models as well. To test for the absorption feature, we multiplied all models by Gaussian absorption profile (GABS) and fit these models to the data. The best-fit parameters are presented in lower rows of Table~\ref{t:x-fit} and the best-fit NSMAX(1200)$\times$GABS model is compared with the data in the right panel of Figure~\ref{fig:x-ray_nsmax}. Besides parameters of the NSMAX model, there are also the line center $E_0$, the full width at half maximum (FWHM), and the equivalent width (EW). In Figure~\ref{fig:nsmaxgabs_triangle}, we also show marginal posterior distributions for the NSMAX(1200)$\times$GABS model parameters. It is seen that the line parameters are well constrained from the fit.

As seen from Table~\ref{t:x-fit}, multiplying each of three models by the Gaussian line with center at about 0.7 keV improves fit statistics. $F$-test value is 19.6 (for model NSMAX 1200) which correspond to the probability of $2.7 \times 10^{-12}$ according to the $F$-distribution. This favors the presence of a line. However it is not correct to use the $F$-distribution to test for a presence of spectral lines \citep[see, e.g.,][]{protassov2002ApJ}. We therefore utilized the method of posterior predictive $p$-values recommended by \citet{protassov2002ApJ}. In brief, we simulated spectra under the null model (model NSMAX in our case) taking parameters from the corresponding posterior distribution, fit them by the null model and an alternative model (in our case, the null model multiplied by GABS), and then computed the $F$-test statistics for each of simulated spectra, whereby constructing the reference distribution for the $F$-test statistics. Comparing the $F$-test value calculated for the data with the reference distribution for the $F$-test statistics based on 5000 data sets simulated under the absorbed NSMAX 1200 model gives $p < 0.0002$ which also favors the presence of an absorption line. The same is true for the models 123100 and 123190.

To set an upper limit on a non-thermal flux of Calvera, we added a power-law (PL) component to the NSMAX(1200)$\times$GABS model and performed spectral fitting. From  resulting joint posterior distribution of the spectral index $\Gamma$ and the normalization of the PL component, we got the distribution of the unabsorbed non-thermal X-ray flux in the 2--8 keV range and then obtained its upper limit, $F_X < 4 \times 10^{-14}$ erg~cm$^{-2}$~s$^{-1}$, as 99.7\% quantile of the distribution. We did not constrain neither $N_{\rm H}$ nor $\Gamma$ during the fits. In this case, the PL component tends to fit the soft part of the spectrum, requiring unrealistically high $N_{\rm H}$ and large $\Gamma$ which is not expected for the pulsars. However, in the 2--8 keV range, the upper limit on the non-thermal flux is dominated by PL models with moderate $\Gamma < 4$ constrained by the hard-energy part of the X-ray spectrum. This approach provides a conservative estimate of the non-thermal limit consistent with that reported by \citet{zane2011MNRAS}, who fixed $\Gamma$ at two.

\section{Discussion}\label{sec:disc}

Deepest up to date optical observations of Calvera allowed us to detect new optical objects in the arcsecond vicinity of the pulsar. The nearest object $E$ is red and locates only 0\farcs67 from the Calvera position. However, precise astrometry showed that the offset significance is about 4.5$\sigma$ thus $E$ can hardly be the Calvera optical counterpart. The GTC data allow us to put upper limits on the fluxes of Calvera in the $g'$ and $i'$ bands (Table~\ref{t:phot}). Assuming a flat spectrum, as it is usually observed for pulsars in the optical band, we can transform the most deep $g'$-band limit into the limit on the unabsorbed flux in the $V$ band, $F_V < 2.5 \times 10^{-17}$ erg~cm$^{-2}$~s$^{-1}$. To account for the interstellar absorption, we use the total Galactic absorption in the direction to Calvera, $A_g = 0.11$ \citep{shlegel1998ApJ}, as a conservative estimate. We also set the upper limit on the non-thermal X-ray flux of Calvera based on the archival X-ray data.

\begin{figure}[t]
  \setlength{\unitlength}{1mm}
  \begin{picture}(160,60)(0,0)
    \put (0.0,0.0)  {\includegraphics[scale=0.4]{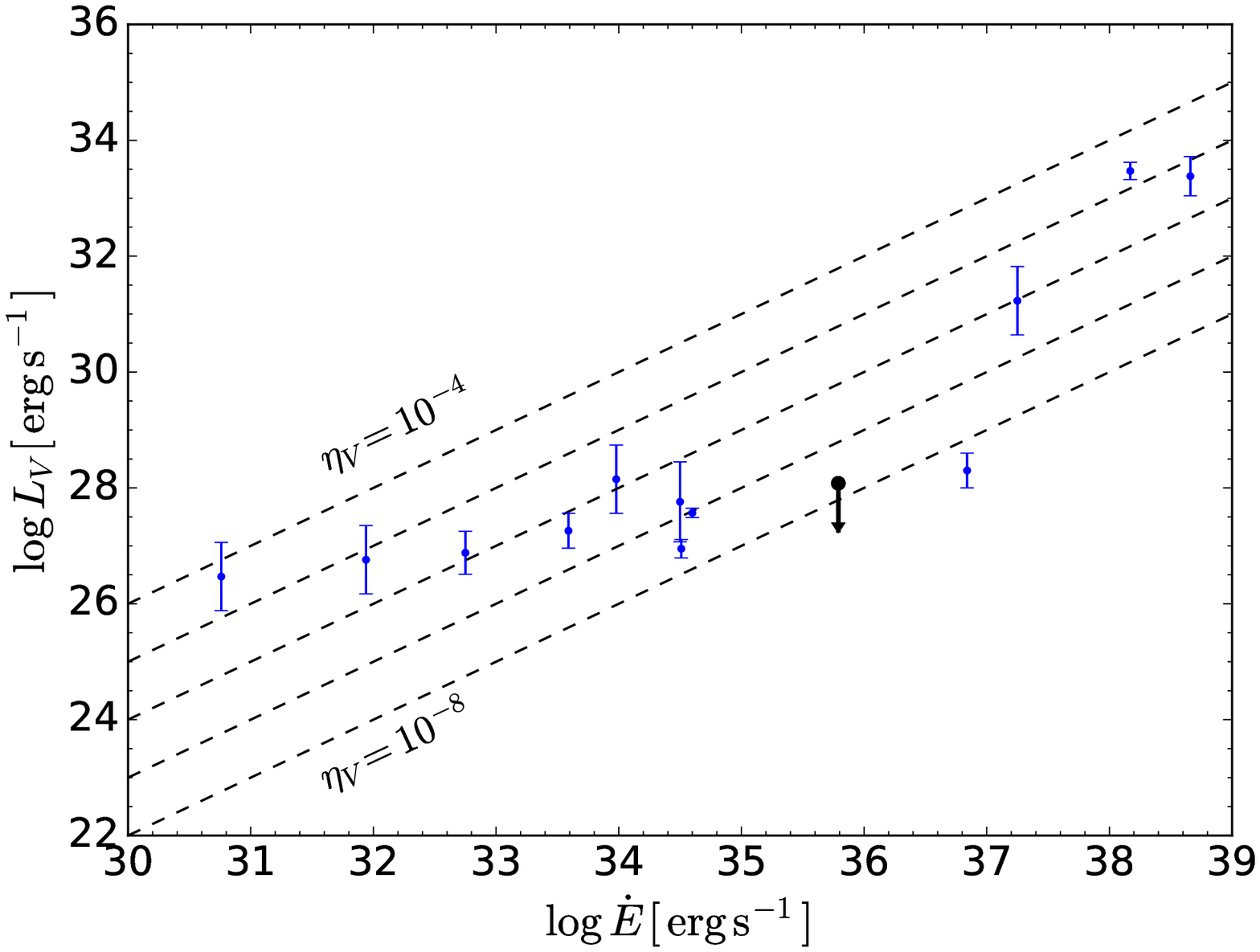}}
    \put (85.0,0.0) {\includegraphics[scale=0.4]{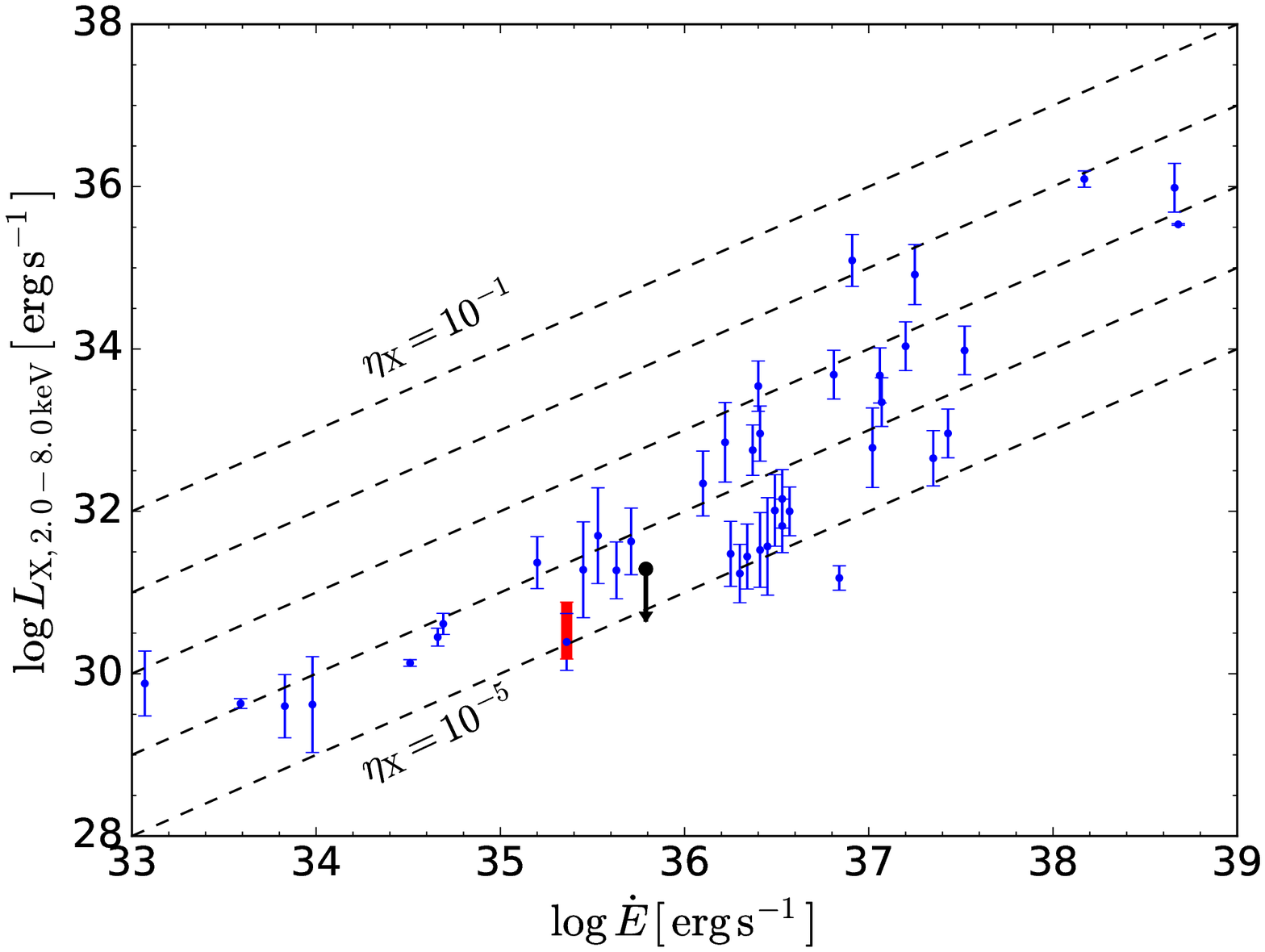}}
  \end{picture}
  \caption{Non-thermal optical and X-ray luminosities vs. $\dot{E}$ for various pulsars. Optical luminosities are in the $V$ band and X-ray luminosities are in the 2--8 keV range. Calvera upper limits for the distance of 2 kpc are shown by black dots with vertical arrows. Dashed lines in both panels correspond to various efficiencies stepped by order of magnitude. The lowest and highest efficiencies are given. The red bar marks the X-ray non-thermal luminosity of PSR~J1740+1000.}
  \label{fig:opt-x-ray_lum}
\end{figure}

The flux upper limits constrain the Calvera optical and X-ray non-thermal luminosities, $L_V < 1.2 \times 10^{28} d_{\rm 2kpc}^{2}$ erg~s$^{-1}$ and $L_{\rm X,2-8 \, keV} < 2 \times 10^{31} d_{\rm 2kpc}^{2}$ erg~s$^{-1}$, where $d_{\rm 2kpc}$ is the distance in units of 2 kpc. In Figure~\ref{fig:opt-x-ray_lum}, we compare them with corresponding luminosities of other pulsars on $L-\dot{E}$ diagrams. The X-ray data are adapted from \citet{kargaltsev2008}\footnote{Their luminosities in 0.5--8.0 keV band were transformed to luminosities in 2.0--8.0 keV band.} and the optical data are taken from \citet{danilenko2013AsAp}. We also add the data on PSR~J1741$-$2054. The X-ray luminosity is taken form \citet{karpova2014ApJ} and the optical one from \citet{mignani2016arXiv}. For both bands, a factor of two uncertainty on the distance is taken into account \citep{karpova2014ApJ,mignani2016arXiv}. Dashed lines correspond to different optical and X-ray efficiencies $\eta_{V} \equiv L_{V}/\dot{E}$ and $\eta_{\rm X} \equiv L_{\rm X}/\dot{E}$. The least efficient pulsar is Vela and other pulsars have optical efficiencies in the range of $10^{-8}$--$10^{-4}$ and X-ray efficiencies in the range of $10^{-5}$--$10^{-1}$. It is seen that Calvera is not an outlier on these diagrams if we assume the distance of 2 kpc or more. It is possible that Calvera is as inefficient in the optical and X-rays as the Vela pulsar.

According to results of X-ray spectral fittings (Table~\ref{t:x-fit}), such a distance implies that an observed thermal emission originates from a significant portion of a NS surface (assuming standard NS radii of $\sim 10-15$~km). The basic argument against this possibility is a relatively large pulsed fraction of the Calvera X-ray emission \citep{halpern2013ApJ}. It was shown by \citet{Page1995} that a NS with a smooth surface temperature map emitting as a blackbody can show no more than 10\% of the pulsed fraction due to gravitational bending effects. However, for magnetic atmospheres the situation is different as the specific flux emerging from the surface has the intrinsic anisotropy. This anisotropy results in the `pencil'+`fan' emission beam shape \citep{zavlin1996AsAp} which is a consequence of the magnetic-field-induced anisotropy of the radiative transfer in atmospheric layers. Therefore, a NS covered with magnetized atmosphere can show pulsations even if its surface temperature is uniform \citep{shibanov1995NYASA}. Interestingly, the calculations given in \citet{shibanov1995NYASA} result in a similar dependence of the pulsed fraction on photon energy as observed for Calvera. Namely, \citet{halpern2013ApJ} found that the pulsed fraction is $\approx 10\%$ at energies below 0.5 keV and $\approx 30\%$ above 0.5 keV, which is in accord with the pulsed fraction calculated for the NS with dipole magnetic field $B = 10^{12}$ G covered by the hydrogen atmosphere \citep{shibanov1995NYASA}.

We found that the X-ray spectrum of Calvera is indeed well fitted by the hydrogen atmosphere model NSMAX with addition of the spectral absorption feature at about $0.7$~keV (Table~\ref{t:x-fit}). The similar conclusion was reached by \citet{zane2011MNRAS} who used the NSA model for the thermal continuum. The atmospheric model 1200 with uniform distribution of temperature over the NS surface and radial magnetic field results in the distance $D\sim 5$ kpc if we assume that X-rays are coming from the bulk of the NS surface. While this model may be suitable for the phase-averaged spectrum, it of course cannot explain pulsations. In order to quantify the pulse shape and to get more relevant model for the phase-averaged spectrum, we need to account for (or fit) the pulsar geometry. In other words, we need to account for the relative position of three vectors -- the magnetic axis, the rotational axis, and the line of sight. For instance, the results of \citet{shibanov1995NYASA} are for the orthogonal rotator, when the rotational axis is perpendicular to both the magnetic axis and the line of sight.

The quantitative fit of the pulsar geometry to the observed phase-dependent spectrum is a good project for the future and is outside the scope of the present paper. Attempts of performing such fits taking into account the atmospheric effects in an approximate way are given in, e.g., \citet{shabaltas2012ApJ} and \citet{Bogdanov2014}. A possible impact of the more realistic consideration is illustrated by the fit results with NSMAX models 123100 and 123190. These models are calculated assuming the dipole surface magnetic field with $B=10^{12}$~G at the magnetic equator and corresponding non-uniform temperature distribution \citep{ho2008ApJS}. The results following from these models can be considered as limiting cases of the real pulsar viewing geometry since the model 123100 assumes that the line of sight coincides with the magnetic moment direction, while in case of the 123190 model, these vectors are orthogonal. The former case corresponds to the situation when hotter polar regions of the star are seen most of the time, while in the latter case, one observes colder equatorial regions. As follows from Table~\ref{t:x-fit}, Calvera can be at $1.5-5$~kpc if covered with magnetized hydrogen atmosphere. The large distance to Calvera implies that $N_{\rm H}$ must be compatible with the total Galactic absorption column density in this direction, $N_{\rm H} = 2.7 \times 10^{20}$ cm$^{-2}$ \citep{karberla2005AsAp}. $N_{\rm H}$ values derived from the 1200 and 123100 fits with the absorption line are  consistent with the total Galactic column density. Spectral fits for the 12390 model formally give twice lower upper limit. However, we checked that fixing $N_{\rm H}$ at $2.7\times 10^{20}$ cm$^{-2}$ does not substantially affect the fit quality and parameter values.

If the distance to Calvera is indeed 1.5~kpc or more, the pulsar is high above the Galactic disc and either it was ejected from there at a high speed or its progenitor was a runaway star. The relatively small proper motion measured by \citet{Halpern2015} invalidates the former possibility. It is interesting, that a similar situation occurs for the radio-pulsar J1740+1000 which resembles Calvera \citep{halpern2013ApJ}. It is also undetected in $\gamma$-rays with the upper limit on the $\gamma$-ray luminosity order of magnitude below those of pulsars with similar $\dot{E}$. The X-ray non-thermal luminosity of J1740+1000 is marked by the red bar in Figure~\ref{fig:opt-x-ray_lum}. It is seen that J1740+1000 is an inefficient X-ray pulsar. The J1740+1000 distance of 1.4~kpc based on dispersion measure (DM) \citep{mclaughlin2002ApJ} implies that it is also high above the Galactic disk, however no significant proper motion was found \citep{halpern2013ApJ}. Thus the runaway-star progenitor is the only possibility for PSR~J1740+1000 (if the DM distance is correct). The runaway-star interpretation for the Calvera progenitor seems plausible as well.

\begin{figure}[t]
  \begin{center}
    \includegraphics[scale=0.60,clip]{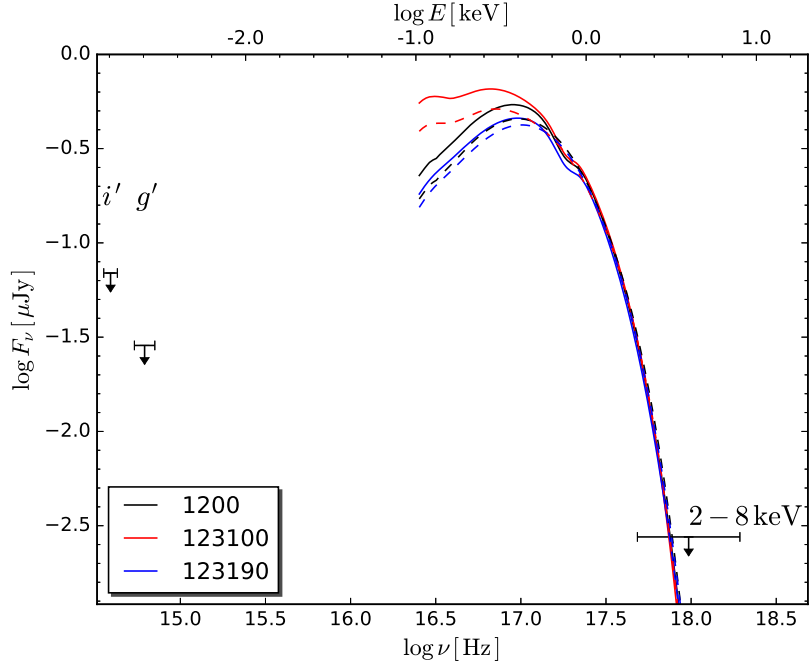}
  \end{center}
  \caption{The best-fit unabsorbed X-ray spectrum of Calvera along with the optical upper limits and the upper limit on the non-thermal X-ray flux. Solid and dashed lines are for the hydrogen atmosphere models NSMAX with and without the absorption feature, respectively.}
  \label{fig:mw_spec}
\end{figure}

The above consideration shows that relatively large distance to Calvera can be in accord with its thermal spectral properties and optical and X-ray non-thermal efficiencies. In this case, Calvera can be an ordinary RPP. The most puzzling thing however is non-detection of the $\gamma$-ray emission with the upper limit on the $\gamma$-ray luminosity $L_{\gamma} < 3.3 \times 10^{32} \, d_{2 \rm kpc}^{2}$ erg~s$^{-1}$ \citep{halpern2013ApJ}. Even if Calvera is at the distance of 5 kpc, it is at the lower edge of corresponding luminosities of other pulsars with similar $\dot{E}$, see figure~5 from \citet{halpern2013ApJ}.

Finally, in Figure~\ref{fig:mw_spec}, we present current update of the multiwavelength appearance of Calvera. The upper limits on the  optical and non-thermal X-ray unabsorbed flux densities shown with arrows are confronted there with the best-fit atmospheric models (Table~\ref{t:x-fit}) extrapolated to the lower energies\footnote{We do not continue atmospheric models in the optical, since the NSMAX calculations are not available there.}. The thermal component is clearly unreachable in the optical with current instruments \citep{zane2011MNRAS} but if Calvera is an RPP, the possible non-thermal component could have been seen. For the middle-aged RPPs detected in the optical and X-rays, the  peak of the NS surface thermal emission usually exceeds the non-thermal background of the pulsar magnetosphere origin however the magnitude of the excess can be different. For instance, for PSR~J1741$-$2054, whose optical counterpart candidate was recently detected \citep{mignani2016arXiv}, this excess is about one order of magnitude. The same situation probably occurs for PSR~J1357$-$6429 \citep[][figure 3]{zuyzin2016MNRAS}. In contrast, for well studied Geminga, PSR B0656+14, and PSR B1055$-$52 thermal peak heights are 2--3 magnitudes higher than the non-thermal level \citep{kargaltsev2005ApJ,mignani2010ApJ,durant2011ApJ}. The relative position of the Calvera optical upper limit with respect to the thermal peak is intermediate between these possibilities encouraging its deeper studies.

\acknowledgments We thank the anonymous referee for useful comments which help us to improve the paper. We also thank Eric Gotthelf, Dima Barsukov, and Serge Balashev for helpful discussion. The work was supported by the Russian Science Foundation, grant 14-12-00316. The scientific results reported in this article are partially based on data obtained from the Chandra Data Archive, observations made by the Chandra X-ray Observatory.

{\it Facilities:} \facility{GTC, \textit{CXO}, \textit{XMM-Newton}, Gemini-N}.

%%%%%%%%%%%%%%%%%%% REFERENCES %%%%%%%%%%%%%%%%%%%%%%%%%%%%%%%%%%%%%%%%%%%%%%%%%%%%%%%%
\bibliographystyle{aasjournal}
\bibliography{ref}

\section*{Non-linearity of the HRC plate}

Usually, the HRC plate is assumed to be linear at least in several arcminutes from the telescope optical axis \citep[e.g,][]{Halpern2015, Beckerman2004}. Since the accurate atrometry uncertainty estimation is crucial for our goals, we directly examined the possible non-linearity of the HRC-I image. That is, we selected archival HRC-I observation of $\sigma$ Orionis (ObsId 2569, PI Wolk) containing sufficient amount of the point-like X-ray sources which have well-defined optical counterparts in the 2MASS catalog. Using IRAF/daophot/ccmap task, we then constructed two astrometric solutions assuming pure linear transformation and the general one. For referencing, we used 22 sources located within 5\farcm5 from the optical axis. The difference between the WSC coordinates given by these solutions was found to be less than 0\farcs05 inside the central 2\arcmin\ detector area. We consider this difference as a systematic error to include in the overall HRC-I astrometric uncertainty budget.

%% the file, uncomment the \plotone, \plottwo, and
%% If you need a layout that cannot be achieved with \plotone or
%% \plottwo, you can invoke the graphicx package directly with the
%% \includegraphics command or use \plotfiddle. For more information,
%% please see the tutorial on "Using Electronic Art with AASTeX" in the
%% documentation section at the AASTeX Web site, http://aastex.aas.org/
%%
%% The examples below also include sample markup for submission of
%% supplemental electronic materials. As always, be sure to check
%% the instructions to authors for the journal you are submitting to
%% for specific submissions guidelines as they vary from

%% This example uses \plotone to include an EPS file scaled to
%% 80% of its natural size with \epsscale. Its caption
%% has been written to indicate that additional figure parts will be
%% available in the electronic journal.

\clearpage

%% Here we use \plottwo to present two versions of the same figure,
%% one in black and white for print the other in RGB color
%% for online presentation. Note that the caption indicates
%% that a color version of the figure will be available online.
%%

\end{document}